\documentclass[longbibliography,aps,prl,twocolumn,superscriptaddress,showpacs]{revtex4-1}
\usepackage{amsthm,amsfonts,amsmath,latexsym,amssymb,color}
\usepackage{amssymb,latexsym,amsfonts,amsmath,amsthm}
\usepackage{hyperref}
\usepackage[inline]{enumitem}
\usepackage{mathtools}
\usepackage{datetime}
\theoremstyle{plain}
\newtheorem*{thm}{Theorem}
\newtheorem*{lem}{Lemma}
\newdateformat{mydate}{\THEDAY\  \monthname[\THEMONTH] \THEYEAR}
\def\ceq{\coloneqq}
\begin{document}

%\preprint{}

\title{Scaling of R\'enyi entanglement entropies of the free Fermi-gas ground state: \\ a rigorous proof}


\author{Hajo Leschke}
\email[]{hajo.leschke@physik.uni-erlangen.de}
%\thanks{}
\affiliation{Institut f\"ur Theoretische Physik, Universit\"at 
Erlangen-N\"urnberg, Staudtstra\ss e 7, 91058 Erlangen, Germany}
\affiliation{Fakult\"at f\"ur Mathematik und Informatik, FernUniversit\"at Hagen, 
Universit\"atsstra\ss e 1, 58097 Hagen, Germany}

\author{Alexander V.~Sobolev}
\email[]{a.sobolev@ucl.ac.uk}
%\thanks{}
\affiliation{\mbox{Department of Mathematics, University College London, Gower Street, London, WC1E 6BT, United~Kingdom}}

\author{Wolfgang Spitzer}
\email[]{wolfgang.spitzer@fernuni-hagen.de}
%\thanks{}
\affiliation{Fakult\"at f\"ur Mathematik und Informatik, FernUniversit\"at Hagen, 
Universit\"atsstra\ss e 1, 58097 Hagen, Germany}


%\date{\today}

\begin{abstract}
In a remarkable paper [Phys.~Rev.~Lett. {\bf 96}, 100503 (2006)], Dimitri Gioev and Israel Klich conjectured an explicit  
formula for the leading asymptotic growth of the spatially bi-partite \mbox{von-Neumann} entanglement 
entropy of non-interacting fermions in multi-dimensional Euclidean space at zero temperature. Based on 
recent progress by one of us (A.\,V.~S.) in semi-classical functional calculus for pseudo-differential operators
with discontinuous symbols, we provide here a complete proof of that formula 
and of its generalization to R\'enyi entropies of all orders $\alpha>0$. The special case $\alpha=1/2$ 
is also known under the name logarithmic negativity and often considered to be a 
particularly useful quantification of entanglement. These formulas, exhibiting a 
``logarithmically enhanced area law'', have been used already in many publications.
\end{abstract}

\pacs{03.65.Ud, 03.67.Mn, 05.30.Fk\\ Appeared, slightly shortened, as Phys. Rev. Lett. {\bf 112}, 160403, 5pp (2014)
 \\\hfill arXiv: 1312.6828 [math-ph] \\ Date of this version: \mydate\today}
% 03.65.Ud: Entanglement and quantum nonlocality
% 03.67.Mn: Entanglement measures, witnesses, and other characterizations 
% 05.30.Fk: Fermion systems and electron gas
% 71.10.Ca: Electron gas, Fermi gas
%\keywords{}

\maketitle


%\paragraph{Introduction.} 
Entanglement is an important property of states of composite quantum systems and has
been established as a key concept of quantum-communication and information theory.
For example, quantum teleportation and quantum computing rely on it \cite{NC,Hay,A,Stol}.
Partially triggered by the above theories, quantifications of entanglement in terms of a suitable 
entropy concept has recently found considerable interest, with an on-going discussion, in 
the theory of quantum many-particle systems \cite{AFOV,ECP,P}. The resulting \emph{entanglement entropy}
 (EE) is not easy to compute --- not even for simple systems in states of thermal 
equilibrium. The reason for that are the quantum correlations between the particles caused 
by their (effective) interaction, which dominate at low temperatures and especially at zero
temperature. However, according to various studies the \emph{spatially} 
bi-partite EE of the (pure) ground state of a ``typical'' quantum many-particle system in
an infinitely extended position space, given either by the $d$-dimensional Euclidean
space $\mathbb R^d$ or the simple cubic lattice $\mathbb Z^d$ ($d=1,2,3,\ldots$) is widely believed to obey
an asymptotic scaling law in the following sense: the entropy (and therefore the 
EE) of such a ground state when spatially reduced to a bounded subregion $\Omega$ of $\mathbb R^d$ or
of $\mathbb Z^d\subset\mathbb R^d$ grows to leading order proportional to the area $L^{d-1} |\partial\Omega|$ of the 
boundary surface $\partial(L\Omega)$ of the scaled region $L\Omega\ceq\{Lq\in\mathbb R^d : q\in\Omega\}$
 as the (dimensionless) scaling parameter $L\ge 1$ tends to infinity, $L\to\infty$.

To our knowledge, this so-called \emph{area law} was first pointed out for free bosonic toy models
\cite{Bom,Sred} in connection with the Bekenstein--Hawking entropy of black holes. Since then 
many interesting results and conjectures were put forward in relation to this area law. For 
example, its simplicity for $d=1$ can be made responsible for the fact that ground states of infinite
quantum spin chains with a spectral gap above their respective ground-state energy may be closely approximated
by finitely correlated (in other words, matrix-product) states \cite{FNW} and are therefore well accessible
to numerically efficient methods like the density-matrix renormalization group \cite{White,Scholl,H}.
We recommend \cite{ECP} for a recent review of such area laws, with an emphasis on rigorous results 
for many-particle systems on the lattice $\mathbb Z^d$.

It is somewhat surprising that an area law is not quite valid for the simple system of a free Fermi gas in $\mathbb R^d$ or $\mathbb Z^d$. Indeed, various studies \cite{Vidal,FHM,K,JK,Wolf,G2,GiKl,BCS,LDYRH,FZ,HLS,Swin1,CMV11,DSY,CMV12,*CMV12b,Swin3,Swin2} have suggested that the EE of its ground state grows (at least as fast) as $L^{d-1}\ln{L}$. Physically, this slight logarithmic enhancement is due to the effective \emph{long-range} correlations of the particles by the Fermi--Dirac statistics, the algebraic statement of Pauli's exclusion principle. Mathematically, the resulting ``sharp'' Fermi surface modifies the asymptotic analysis as $L\to\infty$ in such a way that the extra factor $\ln{L}$ emerges.

\paragraph{The theorem.}
In this Letter, we give a complete proof of this ``logarithmically enhanced area law'' for the free Fermi gas in $\mathbb R^d$ for all $d\ge1$. In order to state our theorem we need some preparations.
\begin{enumerate}[label=(\alph*),labelindent=0\parindent,leftmargin=0\parindent,itemindent=*,noitemsep,nolistsep]
\item The \emph{free Fermi gas} in $\mathbb R^d$ is a well-established model of non-interacting particles obeying Fermi--Dirac statistics. Despite its simplicity, its early spectacular successes in explaining properties of metals and white dwarfs have boosted Quantum Statistical Mechanics \cite{Ba}. We assume the particles to be spin-less and to be characterized by a common translation-invariant one-particle Hamiltonian of the form $\varepsilon(P)$, that is, by some function $\varepsilon:\mathbb R^d\to\mathbb R$ of the canonical momentum operator $P\ceq-\text i\hbar\nabla = -\text i\hbar \,\partial/\partial q$, where $2\pi\hbar>0$ is Planck's constant. We assume $\varepsilon$ to be non-negative and (for simplicity) to be smooth. Then $\varepsilon(P)$ is easy to define as a self-adjoint operator acting on the one-particle Hilbert space $\text L^2(\mathbb R^d)$ of all complex-valued and Lebesgue square-integrable functions $\psi:\mathbb R^d\to\mathbb C$, $q\mapsto \psi(q)$. We also assume that $\varepsilon(p)$ tends to infinity as $|p|\to\infty$. Then the lower-level set $\Gamma\ceq\{p\in\mathbb R^d:\varepsilon(p)\le \varepsilon_F\}$, the \emph{Fermi sea} corresponding to a given finite Fermi energy $\varepsilon_F>0$, is a bounded region in momentum space. More precisely, if $d=1$, then $\Gamma$ is just the union of finitely many pair-wise disjoint bounded intervals. Its boundary, the Fermi surface $\partial\Gamma$, is the set of all endpoints of these intervals, and the ``area'' $|\partial\Gamma|$ of $\partial\Gamma$ is defined as the total number of these points.  If $d\geq 2$, the Fermi sea $\Gamma$ is assumed to be a bounded Lipschitz domain with a piece-wise $\text C^3$-smooth boundary $\partial\Gamma$. For the latter notions see \cite{Sob3}. In the theorem below only these properties of $\Gamma$ are required, since it is based on a lemma which does not explicitly use the fact that $\Gamma$ is ``physically'' a lower-level set of some function $\varepsilon$. The prime example for $\varepsilon$ is given by $\varepsilon(p)=p^2/(2\mathsf m)$, corresponding to the non-relativistic kinetic energy (in the absence of a magnetic field) of a particle with mass $\mathsf m>0$. For a general function $\varepsilon$, the mean particle density $\rho>0$ is related to $\varepsilon_F$, as usual, through the volume of $\Gamma$ by $(2\pi\hbar)^d\rho=|\Gamma|$.
\item The bounded region $\Omega$ in position space $\mathbb R^d$, to which the ground state of the free Fermi gas will be reduced, is assumed to be a bounded Lipschitz domain with a piece-wise $\text C^1$-smooth boundary $\partial\Omega$, if $d\geq2$. If $d=1$, then $\Omega$ is assumed to be the union of finitely many pair-wise disjoint bounded intervals, and the meanings of $\partial\Omega$ and $|\partial\Omega|$ are analogous to those of $\partial\Gamma$ and $|\partial\Gamma|$ given above.
\item For $d\geq 2$ we denote by $\sigma$ and $\tau$ the canonical $(d-1)$-dimensional area measures on the boundary surfaces $\partial\Gamma$ and $\partial\Omega$, respectively. By $m(p)\in\mathbb R^d$ and $n(q)\in\mathbb R^d$ we denote the exterior unit normals at $p\in\partial\Gamma$ and $q\in\partial\Omega$, respectively. 
\end{enumerate}

\begin{thm}
Let $\Gamma\subset\mathbb R^d $ be the Fermi sea of the ground state of the free Fermi gas in $\mathbb R^d$, let this ground state be spatially reduced to $\Omega\subset\mathbb R^d$, and let $S_\alpha(\Gamma,\Omega)$ be its R\'enyi entropy of order $\alpha>0$, see \eqref{def:renyi}, \eqref{lfp}, and \eqref{entropy} below. Then, under the above assumptions on $\Gamma$ and $\Omega$, the following asymptotic formula holds:
\begin{equation}\label{GK-formula}
S_\alpha(\Gamma,L\Omega) \!=\!\frac{1\!+\!\alpha}{24\alpha} J(\partial\Gamma,\partial\Omega) L^{d-1} \ln L + o(L^{d-1}\! \ln L),
\end{equation}
as $L\to\infty$. If $d=1$, the constant $J(\partial\Gamma,\partial\Omega)$ is by definition the product $|\partial\Gamma||\partial\Omega|$ of two positive even numbers. If $d\geq 2$, the constant is given by the double surface-integral
\begin{equation}\label{int2}
J(\partial\Gamma,\partial\Omega) \ceq (2\pi\hbar)^{1-d}\!\!\!\!\int\limits_{\partial \Gamma\times\partial \Omega}\!\!\!\!\!\emph{d}\sigma({p}) \emph{d}\tau({q}) \, \big|m(p)\cdot n(q)\big|\,.
\end{equation} 
\end{thm}

\noindent Before presenting a proof we make a couple of remarks.
\begin{enumerate}[label=(\roman*),labelindent=0\parindent,leftmargin=0\parindent,itemindent=*,noitemsep,nolistsep]
\item To our knowledge, the theorem provides the first rigorous result on the precise leading EE-scaling of a model on \emph{continuous} position space. The $\alpha$-dependence of the leading term in \eqref{GK-formula} reflects the fact that R\'enyi entropies are monotonically decreasing in $\alpha$, see for example \cite{HLS}. The important special case $\alpha=1$ corresponds to the von-Neumann entropy. Since the free Fermi-gas ground state is pure, the case $\alpha=1/2$ is also known under the name \emph{logarithmic negativity} and often considered to be a particularly useful quantification of entanglement \cite{VW,Plenio}. The limiting case $\alpha=\infty$ is also of some interest, because the number $\exp{\big(-S_\infty(\Gamma,\Omega)\big)}$ in the closed unit interval ${[0,1]}$ equals the largest eigenvalue of the quasi-free state operator on the fermionic Fock space $\mathcal F(\text{L}^2(\Omega))$ associated with the spatially reduced ground state, that is, with the ``localized'' Fermi projection \eqref{lfp} below.

\item For a \emph{spherical} Fermi surface, $\partial\Gamma=p_F\,\mathbb S^{d-1}$ with radius $p_F>0$, the integral in \eqref{int2} can be calculated explicitly with the result
\begin{equation}\label{J_explicit}
J(\partial\Gamma,\partial\Omega) = \frac{2}{\big(\frac{d-1}{2}\big)!} \, 
\left(\frac{p_F^2}{4\pi\hbar^2}\right)^{\frac{d-1}{2}} \, |\partial\Omega|\,,
\end{equation}
where $(z)!$ denotes the factorial of $z\in\mathbb C$, that is, Euler's gamma function evaluated at $z+1$. For the theory and applications of surface integrals of this genre and the related \emph{spherical cosine transformation} we refer to \cite{Gard}. In case of the prime example, $\varepsilon(p)=p^2/(2\mathsf m)$, the Fermi momentum $p_F$ is simply related to the Fermi energy by $p_F^2 = 2\mathsf m\,\varepsilon_F$. If one relates $p_F$ in \eqref{J_explicit} to a given (bulk) mean particle density $\rho>0$ via $(2\pi\hbar)^d\rho=|\Gamma|$, one obtains the illuminating formula
\begin{equation}
J(\partial\Gamma,\partial\Omega)=\frac{2}{\big(\frac{d-1}{2}\big)!} \textstyle{\big[\big(\frac{d}{2}\big)!\big]^{\frac{d-1}{d}}}N(\partial\Omega)\,,
\end{equation}
where $N(\partial\Omega)\ceq\rho^{\frac{d-1}{d}}|\partial\Omega|$ is the mean number of particles located on the boundary surface of $\Omega$.

\item Formula~\eqref{GK-formula} has been conjectured by an informal application 
\cite{GiKl,HLS} of the \emph{Widom formula} \eqref{main:eq} below, to the non-smooth function 
$f=h_\alpha$. Here, $h_\alpha: \mathbb R \to[0,\ln{2}]$ is defined for $\alpha>0,\alpha\neq 1$ by 
\begin{equation} \label{def:renyi}
h_\alpha(t) \ceq \frac{1}{1-\alpha} \ln{\big[t^\alpha + (1-t)^\alpha\big]}\,,
\end{equation}
if $t\in[0,1]$ and by $h_\alpha(t) \ceq 0$ if $t\notin[0,1]$. For $\alpha=1$ we set  
$h_1(t) \ceq \lim_{\alpha\to1}h_\alpha(t)$ which equals $-t\ln{t} - (1-t) \ln{(1-t)}$ if  
$t\in\,]0,1[$ and is defined to be $0$ if $t\notin\,]0,1[$.

Formula~\eqref{GK-formula} with $\alpha=1$ has appeared meanwhile in many publications \cite{BCS,LDYRH,FZ,HLS,Swin1,CMV11,DSY} and with general $\alpha>0$ (and $d\ge 2$) besides in \cite{HLS} also in \cite{CMV12,Swin2}.

\item Remarkably, the prefactor in \eqref{GK-formula} with general $\alpha>0$ was found earlier by explicit calculations \cite{JK} for non-interacting fermions on the \emph{one-dimensional lattice} $\mathbb Z$ (with the kinetic-energy operator given by the finite-difference Laplacian and with $\Omega\subset\mathbb Z$ being a single finite block of successive sites).
\end{enumerate}

%%%%%%%%%%%%%%%%%%


\paragraph{Proof of the theorem.}
%Set-Up \cite{HLS}.} 
The proof is based on the Widom formula \eqref{main:eq}, given in the lemma below, and on recent results \cite{Sob2,Sob3} by one of us (A.\,V.~S.). For $d=1$, the proof of the lemma is due to \cite{LW}. For $d\ge 2$ it is contained in \cite{Sob1,Sob3} where the general \emph{Widom conjecture} \cite{Widom_82} is proved \cite{[{A proof of the Widom conjecture in the special case with a quadratic $f$ was previously announced in \cite{GiKl} and given in \cite{HLS}}]leer}. A result of \cite{Sob2} is then used to deal with the non-smoothness of the function $h_\alpha$. By this we eventually justify the informal application of \eqref{main:eq} to $f=h_\alpha$. This is by no means obvious and does not follow from standard approximation arguments. Moreover, a function $f$ being even less smooth than $h_\alpha$ may lead to a scaling behavior considerably different from the one given in \eqref{main:eq}.
%as illustrated by a conjecture due to Fisher and Hartwig \cite{FH,DIK}.

Before we show how to deduce formula \eqref{GK-formula} from the lemma below, we recall that the ground state of the free Fermi gas in $\mathbb R^d$ is the gauge-invariant quasi-free state which is characterized by a one-particle density operator given by the \emph{Fermi projection}
\begin{equation}\label{fp}
\Theta\big(\varepsilon_F\openone-\varepsilon(P)\big)=\chi_\Gamma\big(P)
\end{equation}
acting on $\text L^2(\mathbb R^d)$. Here, $\openone$ denotes the identity operator, $\Theta:\mathbb R\to\{0,1\}$ denotes Heaviside's 
right-continuous unit-step function defined by $\Theta(t)\ceq0$ if $t<0$ and $\Theta(t)\ceq1$ if 
$t\ge 0$, and $\chi_\Gamma$  denotes the indicator function of the Fermi sea $\Gamma$. The 
``localized'' Fermi projection
\begin{equation}\label{lfp}
D(\Gamma,\Omega)\ceq\chi_\Omega(Q)\chi_\Gamma(P)\chi_\Omega(Q)
\end{equation}
then simply characterizes the ground state after spatial reduction to the bounded region $\Omega$ in position space, confer \cite{GiKl,HLS}. Here $Q$ is the usual position (in other words, basic multiplication) operator on $\text L^2(\mathbb R^d)$.

The \emph{local} R\'enyi entropies, that is, the R\'enyi entropies of the spatially reduced ground state may now be defined, confer \cite{GiKl,HLS}, in terms of the functions \eqref{def:renyi} as follows
\begin{equation}\label{entropy}
S_\alpha(\Gamma,\Omega)\ceq\mathrm{Tr}  \,h_\alpha\big(D(\Gamma,\Omega)\big)\,,
\end{equation}
where ``$\mathrm{Tr}$'' refers to the usual trace of operators on $\text L^2(\mathbb R^d)$.
The following chain of (in)equalities
\begin{equation}\label{chain}
0=S_\alpha(\Gamma,\mathbb R^d)< S_\alpha(\Gamma,\mathbb R^d\setminus\Omega)=S_\alpha(\Gamma,\Omega)<\infty
\end{equation}
holds for all bounded $\Omega\subset\mathbb R^d$ of positive volume, $|\Omega|>0$.
The first equality is due to the purity of the ground state, reflected by $\big(D(\Gamma,\mathbb R^d)\big)^2=D(\Gamma,\mathbb R^d)=\chi_\Gamma(P)$. The first inequality is due to the fact that this ground state is not a product state, so that reduction leads to a loss of information, a manifestation of quantum entanglement.
The second equality means that the entropy related to $\Omega$ is the same as the entropy related to its (unbounded) complement in $\mathbb R^d$. This follows from two identities. The first one is the operator identity $h_\alpha\big(\chi_\Gamma(P)\chi_{\mathbb R^d\setminus\Omega}(Q)\chi_\Gamma(P)\big)=h_\alpha\big(\chi_\Gamma(P)\chi_\Omega(Q)\chi_\Gamma(P)\big)$, and the second one is the trace identity $\mathrm{Tr}\,h_\alpha(EFE)=\mathrm{Tr}\,h_\alpha(FEF)$ for any two projections $E$ and $F$ on $\text L^2(\mathbb R^d)$.
The second inequality in \eqref{chain} follows from estimating $h_\alpha(t)$ and the eigenvalues of $D(\Gamma,\Omega)$ from above as done for $\alpha=1$ in \cite{GiKl}. Without the second inequality the question about the asymptotic behavior of $S_\alpha(\Gamma,L\Omega)$ would not make sense. According to \eqref{chain} the \emph{mutual information} $S_\alpha(\Gamma,\Omega)+S_\alpha(\Gamma,\mathbb R^d\setminus\Omega)-S_\alpha(\Gamma,\mathbb R^d)$ equals $2S_\alpha(\Gamma,\Omega)$. Therefore it is justified to identify the local entropy \eqref{entropy} with an EE for any $\alpha>0$.

Next, we want to apply the following lemma due to \cite{LW,Widom_82,Sob1} and due to \cite{Sob3} for 
the extension from smooth to piece-wise smooth boundary surfaces $\partial\Gamma$ and $\partial\Omega$ 
(if $d\ge 2$).

\begin{lem}[Widom formula, a special case]\label{C2:prop}
Let $\Gamma$ and $\Omega$ be bounded subsets of $\mathbb R^d$ with the properties described 
in the preparations (a) and (b) for the theorem. For each $L\ge 1$, let $T_L\ceq D(L\Gamma,\Omega)$
be viewed as a self-adjoint (pseudo-differential) operator on $\emph L^2(\mathbb R^d)$. Finally, 
let $f:\mathbb R \to \mathbb R$ be an infinitely differentiable function ($f\in\text 
C^\infty(\mathbb R)$ for short) with the property $f(0) = 0$. Then the following asymptotic formula 
holds:
\begin{multline}\label{main:eq}
\mathrm{Tr}\, f\big(T_L\big) = f(1)(2\pi\hbar)^{-d}|\Gamma| |\Omega|L^d 
\\
+ I(f) \, J(\partial\Gamma,\partial\Omega)\,L^{d-1} \ln L + o(L^{d-1}\ln L)\,,
\end{multline}
as $L\to\infty$. Here, the linear functional $f\mapsto I(f)$ is defined by the integral
\begin{equation}\label{GA:eq}
I(f) \ceq \frac{1}{4\pi^2}\int_0^1 \!\!\!\emph dt\,\frac{f(t) - t f(1)}{t(1-t)}  \,.
\end{equation} 
\end{lem}
For the applicability of the lemma to the present situation  we first note that $\mathrm{Tr} 
f\big(D(\Gamma,L\Omega)\big)=\mathrm{Tr} f\big(T_L\big)$. This follows from the fact that the 
trace is invariant under the (unitary) dilatation $U_L$ on $\text L^2(\mathbb R^d)$ defined by 
$\big(U_L\psi\big)(q) \ceq L^{d/2} \psi(Lq)$ for all $\psi\in\text L^2(\mathbb R^d)$. 

Given the lemma, for the proof of \eqref{GK-formula} it remains to show how to extend \eqref{main:eq} 
to $f=h_\alpha$, because $h_\alpha\notin \text C^\infty(\mathbb R)$. More generally, we will
show that \eqref{main:eq} can be extended to any function $g:\mathbb R\to\mathbb R$
with the following two properties:
\begin{enumerate}[label=(\roman*)]
\item $g\in\text C^\infty(\mathbb R\setminus \{0,1\})$, that is, $g$ is $\text C^\infty$-smooth
on the doubly pricked real line, $\mathbb R\setminus\{0,1\}$.
\item There exist two constants $C\in\,]0,\infty[$ and $\beta\in\,]0,\infty[$ such that
$|g(t)|\le C\,t^\beta (1-t)^\beta$ for all $t\in\mathbb [0,1]$ (which, in particular, implies $g(0) = g(1) = 0$).
\end{enumerate}
To this end, we are going to mollify $g$ and to control the resulting error. We pick a function 
$w\in\text C^\infty(\mathbb R)$ with $w(\mathbb R)\subseteq[0,1]$, $w(t)=0$ if $|t|>2$, and $w(t)=1$ 
if $|t|\le 1$. Moreover, for each 
$\delta>0$ we define a function $v_\delta\in\text C^\infty(\mathbb R)$ by setting $v_\delta(t)
\ceq w\big(\delta^{-1}t(1-t)\big)$. Since $g(1-v_\delta)\in\text C^\infty(\mathbb R)$ and 
$g(0)(1-v_\delta(0))=0$, the asymptotic formula \eqref{main:eq} holds for 
$f = g(1-v_\delta)$. Observing that this $f$ also fulfills $f(1)=0$, the leading 
$L^d$-term in \eqref{main:eq}, the so-called Weyl term, is seen to vanish and we get
\begin{eqnarray} \label{main term}
\lefteqn{\mathrm{Tr}\, \big[g(T_L)\big(\openone-v_\delta(T_L)\big)\big]}
\\&=& I\big(g(1-v_\delta)\big) \, J(\partial\Gamma,\partial\Omega) \, 
L^{d-1}\ln L + o(L^{d-1}\ln L) \,.\notag
\end{eqnarray} 
Now it remains to show how  $v_\delta$  can be removed from \linebreak\eqref{main term}  in the limit $\delta \downarrow 0$. To this end, we want to show that
\begin{equation}\label{lito:eq}
\mathrm{Tr}\, \big[g(T_L) v_\delta(T_L)\big]  = o_\delta(1) \,L^{d-1}\ln L
\end{equation}
for all $L\ge 2$ with $o_\delta(1)$ vanishing as $\delta\downarrow 0$. By the second property of $g$ 
we have $|g(t)v_\delta(t)|\le C t^\beta(1-t)^\beta v_\delta(t)$. 
Since $t^\nu(1-t)^\nu v_\delta(t)\le (2\delta)^\nu$ for any $\nu>0$, we may choose 
$\gamma\in\,]0, \beta[$ with $\gamma\le 1$, so that $|g(t)v_\delta(t)|\le 
C(2\delta)^{\beta-\gamma} t^\gamma (1-t)^\gamma$. By that we get  
\begin{equation} \label{gamma trace}
\big|\mathrm{Tr}\, \big[g(T_L)v_\delta(T_L)\big]\big| \le  
C(2\delta)^{\beta-\gamma} \mathrm{Tr}\, \big[T_L (\openone - T_L)\big]^\gamma\,.
\end{equation}
In \cite[Corollary 4.7]{Sob2}, the last trace is estimated, by using Schatten--von-Neumann 
(quasi-)norms \cite{Simon,BS},  in such a way that
\begin{equation} \label{est:q-norm}
\mathrm{Tr}\, \big[T_L (\openone - T_L)\big]^\gamma \le C_\gamma\, L^{d-1}\ln L 
\end{equation}
for all $L\ge 2$ with some constant $C_\gamma\!\in\,]0,\infty[$. By combining \eqref{gamma trace} 
and \eqref{est:q-norm} we get 
\begin{equation*}
\big|\mathrm{Tr}\, \big[g(T_L)v_\delta(T_L)\big]\big|
\le C_\gamma C (2\delta)^{\beta-\gamma}\, L^{d-1} \ln{L}\,.
\end{equation*}
Since $\beta-\gamma>0$, the asymptotic behavior \eqref{lito:eq} follows. At the same time we have
\begin{eqnarray} \label{ineq:14}
4\pi^2\,\big|I(g v_\delta)\big|&\le &\int_0^1 \!\!\!\text dt\, 
\frac{|g(t) v_\delta(t)|}{t(1-t)} 
\\
&\le& C (2\delta)^{\beta-\gamma} \int_0^1 \!\!\!\text dt\,  [t(1-t)]^{\gamma-1} \to 0\notag
\end{eqnarray}
as $\delta\downarrow 0$. By combining \eqref{main term}, \eqref{lito:eq}, and \eqref{ineq:14} we
obtain \eqref{main:eq} with $f$ replaced by $g$ and using $g(1)=0$.

Now we return to the function $h_\alpha$. If $\alpha\not=1$, then we choose $\beta = \alpha$, and 
if $\alpha=1$ we may choose any $\beta\in\,]0,1[$. With these choices $h_\alpha$ satisfies $|h_\alpha(t)| \le C \,t^\beta (1-t)^\beta$ for all $t\in[0,1]$ and we arrive at the asymptotic formula \eqref{GK-formula}
by observing $I(h_\alpha) = (1+\alpha)/(24\alpha)$. For the derivation of the last equality see the Appendix.

\paragraph{Concluding remarks.}

We have rigorously derived an explicit formula for the precise asymptotic growth of all R\'enyi (entanglement) entropies of the mixed quantum state, which is obtained from the (pure) ground state of a translation-invariant system of non-interacting fermions in an infinitely extended multi-dimensional continuous position space by reducing, or ``restricting'', the ground state and, hence, the system to a bounded, but linearly growing, subregion of that space. This subregion may be rather general except that its boundary surface, in two or more dimensions, should be piece-wise sufficiently smooth. A similar assumption is needed for the Fermi surface, that is, the boundary of the bounded subregion in momentum space which characterizes the ground state. Our proof of the formula uses a certain trace estimate to extend the recently proved Widom conjecture in asymptotic analysis from smooth functions to a certain class of non-smooth functions. For R\'enyi entropies of orders $\alpha>1$ the required extension is slightly easier to accomplish without the mentioned trace estimate. In any case, such trace estimates will certainly turn out to be useful also for other interesting (spectral) problems in quantum physics and in classical signal theory, where non-smooth functions of integral or pseudo-differential operators naturally show up.

We think that the asymptotic formula \eqref{GK-formula}, conjectured by Gioev and Klich (for $\alpha=1$) and now proved by us, constitutes a rigorous result suitable to serve as a sound standard of comparison for approximate arguments and numerical approaches, when trying to compute entanglement entropies of more complicated many-fermion states. Such states are, for example, the ground states arising from an interaction between the fermions and/or with an external magnetic field.

\vspace{1em}
\noindent APPENDIX: Derivation of $I(h_\alpha) = (1+\alpha)/(24\alpha)$

\vspace{1em}\noindent
For $\alpha\neq 1$ we write $(1-\alpha)h_\alpha(t) = \alpha \ln{t} + \ln{\big[1+((1-t)/t)^\alpha\big]}$. Then, using the change of variables $s\ceq(1-t)/t$, we get
\begin{equation*}
 I(h_\alpha) = \frac{1}{4\pi^2(1-\alpha)} \lim\limits_{x\to\infty}I_\alpha(x)\,,
\end{equation*}
where the function $I_\alpha:{[0,\infty[}\to\mathbb R$ is defined by
\begin{equation*}  I_\alpha(x) \ceq \int_0^x \frac{\text d s}{s}\,  \big[-\alpha\ln{(1+s)} + \ln{(1+s^\alpha)}\big]\,.
\end{equation*}
For this function we have
\begin{align*}  
I_\alpha(x)&=-\alpha\int_0^x \frac{\text d s}{s}\, \ln{(1+s)} + {\textstyle{\frac1\alpha}} \int_0^{x^\alpha} \frac{\text d r}{r}\, \ln{(1+r)}\\
&=\alpha \,\mathrm{Li}(1+x) - {\textstyle{\frac1\alpha}} \,\mathrm{Li}(1+x^\alpha)\\
&=\alpha \Big[\mathrm{Li}(1+x) + {\textstyle{\frac12}} \big(\ln{(1+x)}\big)^2\Big]\\
&\quad-{\textstyle{\frac1\alpha}} \Big[\mathrm{Li}(1+x^\alpha) + {\textstyle{\frac12}} \big(\ln{(1+x^\alpha)}\big)^2\Big]\\
&\quad-{\textstyle{\frac\alpha2}} \big(\ln{(1+x)}\big)^2 + {\textstyle{\frac1{2\alpha}}} \big(\ln{(1+x^\alpha)}\big)^2\,,
\end{align*}
where $\mathrm{Li}$ denotes Euler's di-logarithm. Now we observe that the last sum tends to zero as $x$ tends to
infinity and that
\begin{equation*}
 \lim_{y\to\infty} \Big[\mathrm{Li}(y) + {\textstyle{\frac12}} (\ln{y})^2\Big] = -\frac{\pi^2}{6}\,.
\end{equation*}
Hence,
\begin{equation*} 
I(h_\alpha) = \frac1{4\pi^2(1-\alpha)} \,\Big(\frac{1}{\alpha} - \alpha\Big) \frac{\pi^2}6 = \frac{1+\alpha}{24\alpha}\,.
\end{equation*}
This equality remains valid in the limit $\alpha\to 1$. In fact, $I(h_1)$ can be calculated directly (and more easily) by using the Mercator--Taylor series of the natural logarithm.
%\vspace{1em}
%\paragraph{Conclusion:}
\bibliography{biblio}

\begin{thebibliography}{43}%
\makeatletter
\providecommand \@ifxundefined [1]{%
 \@ifx{#1\undefined}
}%
\providecommand \@ifnum [1]{%
 \ifnum #1\expandafter \@firstoftwo
 \else \expandafter \@secondoftwo
 \fi
}%
\providecommand \@ifx [1]{%
 \ifx #1\expandafter \@firstoftwo
 \else \expandafter \@secondoftwo
 \fi
}%
\providecommand \natexlab [1]{#1}%
\providecommand \enquote  [1]{``#1''}%
\providecommand \bibnamefont  [1]{#1}%
\providecommand \bibfnamefont [1]{#1}%
\providecommand \citenamefont [1]{#1}%
\providecommand \href@noop [0]{\@secondoftwo}%
\providecommand \href [0]{\begingroup \@sanitize@url \@href}%
\providecommand \@href[1]{\@@startlink{#1}\@@href}%
\providecommand \@@href[1]{\endgroup#1\@@endlink}%
\providecommand \@sanitize@url [0]{\catcode `\\12\catcode `\$12\catcode
  `\&12\catcode `\#12\catcode `\^12\catcode `\_12\catcode `\%12\relax}%
\providecommand \@@startlink[1]{}%
\providecommand \@@endlink[0]{}%
\providecommand \url  [0]{\begingroup\@sanitize@url \@url }%
\providecommand \@url [1]{\endgroup\@href {#1}{\urlprefix }}%
\providecommand \urlprefix  [0]{URL }%
\providecommand \Eprint [0]{\href }%
\providecommand \doibase [0]{http://dx.doi.org/}%
\providecommand \selectlanguage [0]{\@gobble}%
\providecommand \bibinfo  [0]{\@secondoftwo}%
\providecommand \bibfield  [0]{\@secondoftwo}%
\providecommand \translation [1]{[#1]}%
\providecommand \BibitemOpen [0]{}%
\providecommand \bibitemStop [0]{}%
\providecommand \bibitemNoStop [0]{.\EOS\space}%
\providecommand \EOS [0]{\spacefactor3000\relax}%
\providecommand \BibitemShut  [1]{\csname bibitem#1\endcsname}%
\let\auto@bib@innerbib\@empty
%</preamble>
\bibitem [{\citenamefont {Nielsen}\ and\ \citenamefont {Chuang}(2000)}]{NC}%
  \BibitemOpen
  \bibfield  {author} {\bibinfo {author} {\bibfnamefont {M.~A.}\ \bibnamefont
  {Nielsen}}\ and\ \bibinfo {author} {\bibfnamefont {I.~L.}\ \bibnamefont
  {Chuang}},\ }\href@noop {} {\emph {\bibinfo {title} {Quantum Computation and
  Quantum Information}}}\ (\bibinfo  {publisher} {Cambridge University Press},\
  \bibinfo {address} {Cambridge},\ \bibinfo {year} {2000})\BibitemShut
  {NoStop}%
\bibitem [{\citenamefont {Hayashi}(2006)}]{Hay}%
  \BibitemOpen
  \bibfield  {author} {\bibinfo {author} {\bibfnamefont {M.}~\bibnamefont
  {Hayashi}},\ }\href@noop {} {\emph {\bibinfo {title} {Quantum Information: An
  Introduction}}}\ (\bibinfo  {publisher} {Springer-Verlag},\ \bibinfo {year}
  {2006})\BibitemShut {NoStop}%
\bibitem [{\citenamefont {Audretsch}(2007)}]{A}%
  \BibitemOpen
  \bibfield  {author} {\bibinfo {author} {\bibfnamefont {J.}~\bibnamefont
  {Audretsch}},\ }\href@noop {} {\emph {\bibinfo {title} {Entangled Systems}}}\
  (\bibinfo  {publisher} {Wiley-VCH, Berlin},\ \bibinfo {year}
  {2007})\BibitemShut {NoStop}%
\bibitem [{\citenamefont {Stolze}\ and\ \citenamefont {Suter}(2008)}]{Stol}%
  \BibitemOpen
  \bibfield  {author} {\bibinfo {author} {\bibfnamefont {J.}~\bibnamefont
  {Stolze}}\ and\ \bibinfo {author} {\bibfnamefont {D.}~\bibnamefont {Suter}},\
  }\href@noop {} {\emph {\bibinfo {title} {Quantum Computing: A Short Course
  from Theory to Experiment, 2nd edition.}}}\ (\bibinfo  {publisher}
  {Wiley-VCH},\ \bibinfo {address} {Weinheim},\ \bibinfo {year}
  {2008})\BibitemShut {NoStop}%
\bibitem [{\citenamefont {Amico}\ \emph {et~al.}(2008)\citenamefont {Amico},
  \citenamefont {Fazio}, \citenamefont {Osterloh},\ and\ \citenamefont
  {Vedral}}]{AFOV}%
  \BibitemOpen
  \bibfield  {author} {\bibinfo {author} {\bibfnamefont {L.}~\bibnamefont
  {Amico}}, \bibinfo {author} {\bibfnamefont {R.}~\bibnamefont {Fazio}},
  \bibinfo {author} {\bibfnamefont {A.}~\bibnamefont {Osterloh}}, \ and\
  \bibinfo {author} {\bibfnamefont {V.}~\bibnamefont {Vedral}},\ }\bibfield
  {title} {\enquote {\bibinfo {title} {Entanglement in many-body systems},}\
  }\href@noop {} {\bibfield  {journal} {\bibinfo  {journal} {Rev. Mod. Phys.}\
  }\textbf {\bibinfo {volume} {80}},\ \bibinfo {pages} {517--576} (\bibinfo
  {year} {2008})}\BibitemShut {NoStop}%
\bibitem [{\citenamefont {Eisert}\ \emph {et~al.}(2010)\citenamefont {Eisert},
  \citenamefont {Cramer},\ and\ \citenamefont {Plenio}}]{ECP}%
  \BibitemOpen
  \bibfield  {author} {\bibinfo {author} {\bibfnamefont {J.}~\bibnamefont
  {Eisert}}, \bibinfo {author} {\bibfnamefont {M.}~\bibnamefont {Cramer}}, \
  and\ \bibinfo {author} {\bibfnamefont {M.~B.}\ \bibnamefont {Plenio}},\
  }\bibfield  {title} {\enquote {\bibinfo {title} {Area laws for the
  entanglement entropy -- a review},}\ }\href@noop {} {\bibfield  {journal}
  {\bibinfo  {journal} {Rev. Mod. Phys.}\ }\textbf {\bibinfo {volume} {82}},\
  \bibinfo {pages} {277--306} (\bibinfo {year} {2010})}\BibitemShut {NoStop}%
\bibitem [{\citenamefont {Peschel}(2012)}]{P}%
  \BibitemOpen
  \bibfield  {author} {\bibinfo {author} {\bibfnamefont {I.}~\bibnamefont
  {Peschel}},\ }\bibfield  {title} {\enquote {\bibinfo {title} {Entanglement in
  solvable many-particle models},}\ }\href@noop {} {\bibfield  {journal}
  {\bibinfo  {journal} {Braz. J. Phys.}\ }\textbf {\bibinfo {volume} {42}},\
  \bibinfo {pages} {267--291} (\bibinfo {year} {2012})}\BibitemShut {NoStop}%
\bibitem [{\citenamefont {Bombelli}\ \emph {et~al.}(1986)\citenamefont
  {Bombelli}, \citenamefont {Koul}, \citenamefont {Lee},\ and\ \citenamefont
  {Sorkin}}]{Bom}%
  \BibitemOpen
  \bibfield  {author} {\bibinfo {author} {\bibfnamefont {L.}~\bibnamefont
  {Bombelli}}, \bibinfo {author} {\bibfnamefont {R.~K.}\ \bibnamefont {Koul}},
  \bibinfo {author} {\bibfnamefont {J.-H.}\ \bibnamefont {Lee}}, \ and\
  \bibinfo {author} {\bibfnamefont {R.~D.}\ \bibnamefont {Sorkin}},\ }\bibfield
   {title} {\enquote {\bibinfo {title} {A quantum source of entropy for black
  holes},}\ }\href@noop {} {\bibfield  {journal} {\bibinfo  {journal} {Phys.
  Rev. D}\ }\textbf {\bibinfo {volume} {34}},\ \bibinfo {pages} {373--383}
  (\bibinfo {year} {1986})}\BibitemShut {NoStop}%
\bibitem [{\citenamefont {Srednicki}(1993)}]{Sred}%
  \BibitemOpen
  \bibfield  {author} {\bibinfo {author} {\bibfnamefont {M.}~\bibnamefont
  {Srednicki}},\ }\bibfield  {title} {\enquote {\bibinfo {title} {Entropy and
  area},}\ }\href@noop {} {\bibfield  {journal} {\bibinfo  {journal} {Phys.
  Rev. Lett.}\ }\textbf {\bibinfo {volume} {71}},\ \bibinfo {pages} {666--669}
  (\bibinfo {year} {1993})}\BibitemShut {NoStop}%
\bibitem [{\citenamefont {Fannes}\ \emph {et~al.}(1992)\citenamefont {Fannes},
  \citenamefont {Nachtergaele},\ and\ \citenamefont {Werner}}]{FNW}%
  \BibitemOpen
  \bibfield  {author} {\bibinfo {author} {\bibfnamefont {M.}~\bibnamefont
  {Fannes}}, \bibinfo {author} {\bibfnamefont {B.}~\bibnamefont
  {Nachtergaele}}, \ and\ \bibinfo {author} {\bibfnamefont {R.~F.}\
  \bibnamefont {Werner}},\ }\bibfield  {title} {\enquote {\bibinfo {title}
  {Finitely correlated states on quantum spin chains},}\ }\href@noop {}
  {\bibfield  {journal} {\bibinfo  {journal} {Commun. Math. Phys.}\ }\textbf
  {\bibinfo {volume} {144}},\ \bibinfo {pages} {443--490} (\bibinfo {year}
  {1992})}\BibitemShut {NoStop}%
\bibitem [{\citenamefont {White}(1992)}]{White}%
  \BibitemOpen
  \bibfield  {author} {\bibinfo {author} {\bibfnamefont {S.~R.}\ \bibnamefont
  {White}},\ }\bibfield  {title} {\enquote {\bibinfo {title} {Density matrix
  formulation for quantum renormalization groups},}\ }\href@noop {} {\bibfield
  {journal} {\bibinfo  {journal} {Phys. Rev. Lett.}\ }\textbf {\bibinfo
  {volume} {69}},\ \bibinfo {pages} {2863--2866} (\bibinfo {year}
  {1992})}\BibitemShut {NoStop}%
\bibitem [{\citenamefont {Schollw{\"o}ck}(2005)}]{Scholl}%
  \BibitemOpen
  \bibfield  {author} {\bibinfo {author} {\bibfnamefont {U.}~\bibnamefont
  {Schollw{\"o}ck}},\ }\bibfield  {title} {\enquote {\bibinfo {title} {The
  density-matrix renormalization group},}\ }\href@noop {} {\bibfield  {journal}
  {\bibinfo  {journal} {Rev. Mod. Phys.}\ }\textbf {\bibinfo {volume} {77}},\
  \bibinfo {pages} {259--315} (\bibinfo {year} {2005})}\BibitemShut {NoStop}%
\bibitem [{\citenamefont {Hastings}(2007)}]{H}%
  \BibitemOpen
  \bibfield  {author} {\bibinfo {author} {\bibfnamefont {M.~B.}\ \bibnamefont
  {Hastings}},\ }\bibfield  {title} {\enquote {\bibinfo {title} {Entropy and
  entanglement in quantum ground states},}\ }\href@noop {} {\bibfield
  {journal} {\bibinfo  {journal} {Phys. Rev. B}\ }\textbf {\bibinfo {volume}
  {76}},\ \bibinfo {pages} {035114} (\bibinfo {year} {2007})}\BibitemShut
  {NoStop}%
\bibitem [{\citenamefont {Vidal}\ \emph {et~al.}(2003)\citenamefont {Vidal},
  \citenamefont {Latorre}, \citenamefont {Rico},\ and\ \citenamefont
  {Kitaev}}]{Vidal}%
  \BibitemOpen
  \bibfield  {author} {\bibinfo {author} {\bibfnamefont {G.}~\bibnamefont
  {Vidal}}, \bibinfo {author} {\bibfnamefont {J.~I.}\ \bibnamefont {Latorre}},
  \bibinfo {author} {\bibfnamefont {E.}~\bibnamefont {Rico}}, \ and\ \bibinfo
  {author} {\bibfnamefont {A.}~\bibnamefont {Kitaev}},\ }\bibfield  {title}
  {\enquote {\bibinfo {title} {Entanglement in quantum critical phenomena},}\
  }\href@noop {} {\bibfield  {journal} {\bibinfo  {journal} {Phys. Rev. Lett.}\
  }\textbf {\bibinfo {volume} {90}},\ \bibinfo {pages} {227902} (\bibinfo
  {year} {2003})}\BibitemShut {NoStop}%
\bibitem [{\citenamefont {Fannes}\ \emph {et~al.}(2003)\citenamefont {Fannes},
  \citenamefont {Haegeman},\ and\ \citenamefont {Mosonyi}}]{FHM}%
  \BibitemOpen
  \bibfield  {author} {\bibinfo {author} {\bibfnamefont {M.}~\bibnamefont
  {Fannes}}, \bibinfo {author} {\bibfnamefont {B.}~\bibnamefont {Haegeman}}, \
  and\ \bibinfo {author} {\bibfnamefont {M.}~\bibnamefont {Mosonyi}},\
  }\bibfield  {title} {\enquote {\bibinfo {title} {Entropy growth of
  shift-invariant states on a quantum spin chain},}\ }\href@noop {} {\bibfield
  {journal} {\bibinfo  {journal} {J. Math. Phys. (N.\,Y.)}\ }\textbf {\bibinfo
  {volume} {44}},\ \bibinfo {pages} {6005--6009} (\bibinfo {year}
  {2003})}\BibitemShut {NoStop}%
\bibitem [{\citenamefont {Korepin}(2004)}]{K}%
  \BibitemOpen
  \bibfield  {author} {\bibinfo {author} {\bibfnamefont {V.~E.}\ \bibnamefont
  {Korepin}},\ }\bibfield  {title} {\enquote {\bibinfo {title} {Universality of
  entropy scaling in one dimensional gapless models},}\ }\href@noop {}
  {\bibfield  {journal} {\bibinfo  {journal} {Phys. Rev. Lett.}\ }\textbf
  {\bibinfo {volume} {92}},\ \bibinfo {pages} {096402} (\bibinfo {year}
  {2004})},\ \Eprint {http://arxiv.org/abs/and arXiv: cond-mat 03110564v4
  (2005)} {and arXiv: cond-mat 03110564v4 (2005)} \BibitemShut {NoStop}%
\bibitem [{\citenamefont {Jin}\ and\ \citenamefont {Korepin}(2004)}]{JK}%
  \BibitemOpen
  \bibfield  {author} {\bibinfo {author} {\bibfnamefont {B.-Q.}\ \bibnamefont
  {Jin}}\ and\ \bibinfo {author} {\bibfnamefont {V.~E.}\ \bibnamefont
  {Korepin}},\ }\bibfield  {title} {\enquote {\bibinfo {title} {Quantum spin
  chain, {Toeplitz} determinants and the {Fisher--Hartwig} conjecture},}\
  }\href@noop {} {\bibfield  {journal} {\bibinfo  {journal} {J. Stat. Phys.}\
  }\textbf {\bibinfo {volume} {116}},\ \bibinfo {pages} {79--95} (\bibinfo
  {year} {2004})}\BibitemShut {NoStop}%
\bibitem [{\citenamefont {Wolf}(2006)}]{Wolf}%
  \BibitemOpen
  \bibfield  {author} {\bibinfo {author} {\bibfnamefont {M.~M.}\ \bibnamefont
  {Wolf}},\ }\bibfield  {title} {\enquote {\bibinfo {title} {Violation of the
  entropic area law for fermions},}\ }\href@noop {} {\bibfield  {journal}
  {\bibinfo  {journal} {Phys. Rev. Lett.}\ }\textbf {\bibinfo {volume} {96}},\
  \bibinfo {pages} {010404} (\bibinfo {year} {2006})}\BibitemShut {NoStop}%
\bibitem [{\citenamefont {Gioev}(2006)}]{G2}%
  \BibitemOpen
  \bibfield  {author} {\bibinfo {author} {\bibfnamefont {D.}~\bibnamefont
  {Gioev}},\ }\bibfield  {title} {\enquote {\bibinfo {title} {Szeg{\H o} limit
  theorem for operators with discontinuous symbols and applications to
  entanglement entropy},}\ }\href@noop {} {\bibfield  {journal} {\bibinfo
  {journal} {Int. Math. Res. Notices}\ }\textbf {\bibinfo {volume} {2006}},\
  \bibinfo {pages} {95181} (\bibinfo {year} {2006})},\ \Eprint
  {http://arxiv.org/abs/or: arXiv: 0212215v4 [math.FA] (2006)} {or: arXiv:
  0212215v4 [math.FA] (2006)} \BibitemShut {NoStop}%
\bibitem [{\citenamefont {Gioev}\ and\ \citenamefont {Klich}(2006)}]{GiKl}%
  \BibitemOpen
  \bibfield  {author} {\bibinfo {author} {\bibfnamefont {D.}~\bibnamefont
  {Gioev}}\ and\ \bibinfo {author} {\bibfnamefont {I.}~\bibnamefont {Klich}},\
  }\bibfield  {title} {\enquote {\bibinfo {title} {Entanglement entropy of
  fermions in any dimension and the {Widom} conjecture},}\ }\href@noop {}
  {\bibfield  {journal} {\bibinfo  {journal} {Phys. Rev. Lett.}\ }\textbf
  {\bibinfo {volume} {96}},\ \bibinfo {pages} {100503} (\bibinfo {year}
  {2006})}\BibitemShut {NoStop}%
\bibitem [{\citenamefont {Barthel}\ \emph {et~al.}(2006)\citenamefont
  {Barthel}, \citenamefont {Chung},\ and\ \citenamefont {Schollw\"ock}}]{BCS}%
  \BibitemOpen
  \bibfield  {author} {\bibinfo {author} {\bibfnamefont {T.}~\bibnamefont
  {Barthel}}, \bibinfo {author} {\bibfnamefont {M.-C.}\ \bibnamefont {Chung}},
  \ and\ \bibinfo {author} {\bibfnamefont {U.}~\bibnamefont {Schollw\"ock}},\
  }\bibfield  {title} {\enquote {\bibinfo {title} {Entanglement scaling in
  critical two-dimensional fermionic and bosonic systems},}\ }\href@noop {}
  {\bibfield  {journal} {\bibinfo  {journal} {Phys.~Rev. A}\ }\textbf {\bibinfo
  {volume} {74}},\ \bibinfo {pages} {022329} (\bibinfo {year}
  {2006})}\BibitemShut {NoStop}%
\bibitem [{\citenamefont {Li}\ \emph {et~al.}(2006)\citenamefont {Li},
  \citenamefont {Ding}, \citenamefont {Yu}, \citenamefont {Roscilde},\ and\
  \citenamefont {Haas}}]{LDYRH}%
  \BibitemOpen
  \bibfield  {author} {\bibinfo {author} {\bibfnamefont {W.}~\bibnamefont
  {Li}}, \bibinfo {author} {\bibfnamefont {L.}~\bibnamefont {Ding}}, \bibinfo
  {author} {\bibfnamefont {R.}~\bibnamefont {Yu}}, \bibinfo {author}
  {\bibfnamefont {T.}~\bibnamefont {Roscilde}}, \ and\ \bibinfo {author}
  {\bibfnamefont {S.}~\bibnamefont {Haas}},\ }\bibfield  {title} {\enquote
  {\bibinfo {title} {Scaling behavior of entanglement in two and
  three-dimensional free fermions},}\ }\href@noop {} {\bibfield  {journal}
  {\bibinfo  {journal} {Phys. Rev. B}\ }\textbf {\bibinfo {volume} {74}},\
  \bibinfo {pages} {073103} (\bibinfo {year} {2006})}\BibitemShut {NoStop}%
\bibitem [{\citenamefont {Farkas}\ and\ \citenamefont
  {Zimbor\'{a}s}(2007)}]{FZ}%
  \BibitemOpen
  \bibfield  {author} {\bibinfo {author} {\bibfnamefont {S.}~\bibnamefont
  {Farkas}}\ and\ \bibinfo {author} {\bibfnamefont {Z.}~\bibnamefont
  {Zimbor\'{a}s}},\ }\bibfield  {title} {\enquote {\bibinfo {title} {The {von
  Neumann} entropy asymptotics in multidimensional fermionic systems},}\
  }\href@noop {} {\bibfield  {journal} {\bibinfo  {journal} {J. Math. Phys.
  (N.\,Y.)}\ }\textbf {\bibinfo {volume} {48}},\ \bibinfo {pages} {102110}
  (\bibinfo {year} {2007})},\ \Eprint {http://arxiv.org/abs/and arXiv:
  0706.1805v2 [math-ph] (2011)} {and arXiv: 0706.1805v2 [math-ph] (2011)}
  \BibitemShut {NoStop}%
\bibitem [{\citenamefont {Helling}\ \emph {et~al.}(2011)\citenamefont
  {Helling}, \citenamefont {Leschke},\ and\ \citenamefont {Spitzer}}]{HLS}%
  \BibitemOpen
  \bibfield  {author} {\bibinfo {author} {\bibfnamefont {R.}~\bibnamefont
  {Helling}}, \bibinfo {author} {\bibfnamefont {H.}~\bibnamefont {Leschke}}, \
  and\ \bibinfo {author} {\bibfnamefont {W.}~\bibnamefont {Spitzer}},\
  }\bibfield  {title} {\enquote {\bibinfo {title} {A special case of a
  conjecture by {Widom} with implications to fermionic entanglement entropy},}\
  }\href@noop {} {\bibfield  {journal} {\bibinfo  {journal} {Int. Math. Res.
  Notices}\ }\textbf {\bibinfo {volume} {2011}},\ \bibinfo {pages} {1451--1482}
  (\bibinfo {year} {2011})},\ \Eprint {http://arxiv.org/abs/or arXiv:
  0906.4946v2 [math-ph] (2010)} {or arXiv: 0906.4946v2 [math-ph] (2010)}
  \BibitemShut {NoStop}%
\bibitem [{\citenamefont {Swingle}(2010)}]{Swin1}%
  \BibitemOpen
  \bibfield  {author} {\bibinfo {author} {\bibfnamefont {B.}~\bibnamefont
  {Swingle}},\ }\bibfield  {title} {\enquote {\bibinfo {title} {Entanglement
  entropy and the {Fermi} surface},}\ }\href@noop {} {\bibfield  {journal}
  {\bibinfo  {journal} {Phys. Rev. Lett.}\ }\textbf {\bibinfo {volume} {105}},\
  \bibinfo {pages} {050502} (\bibinfo {year} {2010})}\BibitemShut {NoStop}%
\bibitem [{\citenamefont {Calabrese}\ \emph {et~al.}(2011)\citenamefont
  {Calabrese}, \citenamefont {Mintchev},\ and\ \citenamefont {Vicari}}]{CMV11}%
  \BibitemOpen
  \bibfield  {author} {\bibinfo {author} {\bibfnamefont {P.}~\bibnamefont
  {Calabrese}}, \bibinfo {author} {\bibfnamefont {M.}~\bibnamefont {Mintchev}},
  \ and\ \bibinfo {author} {\bibfnamefont {E.}~\bibnamefont {Vicari}},\
  }\bibfield  {title} {\enquote {\bibinfo {title} {Entanglement entropy of
  one-dimensional gases},}\ }\href@noop {} {\bibfield  {journal} {\bibinfo
  {journal} {Phys. Rev. Lett.}\ }\textbf {\bibinfo {volume} {107}},\ \bibinfo
  {pages} {020601} (\bibinfo {year} {2011})}\BibitemShut {NoStop}%
\bibitem [{\citenamefont {Ding}\ \emph {et~al.}(2012)\citenamefont {Ding},
  \citenamefont {Seidel},\ and\ \citenamefont {Yang}}]{DSY}%
  \BibitemOpen
  \bibfield  {author} {\bibinfo {author} {\bibfnamefont {W.}~\bibnamefont
  {Ding}}, \bibinfo {author} {\bibfnamefont {A.}~\bibnamefont {Seidel}}, \ and\
  \bibinfo {author} {\bibfnamefont {K.}~\bibnamefont {Yang}},\ }\bibfield
  {title} {\enquote {\bibinfo {title} {Entanglement entropy of {Fermi} liquids
  via multidimensional bosonization},}\ }\href@noop {} {\bibfield  {journal}
  {\bibinfo  {journal} {Phys. Rev. X}\ }\textbf {\bibinfo {volume} {2}},\
  \bibinfo {pages} {011012} (\bibinfo {year} {2012})}\BibitemShut {NoStop}%
\bibitem [{\citenamefont {Calabrese}\ \emph
  {et~al.}(2012{\natexlab{a}})\citenamefont {Calabrese}, \citenamefont
  {Mintchev},\ and\ \citenamefont {Vicari}}]{CMV12}%
  \BibitemOpen
  \bibfield  {author} {\bibinfo {author} {\bibfnamefont {P.}~\bibnamefont
  {Calabrese}}, \bibinfo {author} {\bibfnamefont {M.}~\bibnamefont {Mintchev}},
  \ and\ \bibinfo {author} {\bibfnamefont {E.}~\bibnamefont {Vicari}},\
  }\bibfield  {title} {\enquote {\bibinfo {title} {Entanglement entropies in
  free-fermion gases for arbitrary dimension},}\ }\href@noop {} {\bibfield
  {journal} {\bibinfo  {journal} {Europhys. Lett.}\ }\textbf {\bibinfo {volume}
  {97}},\ \bibinfo {pages} {20009} (\bibinfo {year}
  {2012}{\natexlab{a}})}\BibitemShut {NoStop}%
\bibitem [{\citenamefont {Calabrese}\ \emph
  {et~al.}(2012{\natexlab{b}})\citenamefont {Calabrese}, \citenamefont
  {Mintchev},\ and\ \citenamefont {Vicari}}]{CMV12b}%
  \BibitemOpen
  \bibfield  {author} {\bibinfo {author} {\bibfnamefont {P.}~\bibnamefont
  {Calabrese}}, \bibinfo {author} {\bibfnamefont {M.}~\bibnamefont {Mintchev}},
  \ and\ \bibinfo {author} {\bibfnamefont {E.}~\bibnamefont {Vicari}},\
  }\bibfield  {title} {\enquote {\bibinfo {title} {Exact relations between
  particle fluctuations and entanglement in {Fermi} gases},}\ }\href@noop {}
  {\bibfield  {journal} {\bibinfo  {journal} {Europhys. Lett.}\ }\textbf
  {\bibinfo {volume} {98}},\ \bibinfo {pages} {20003} (\bibinfo {year}
  {2012}{\natexlab{b}})}\BibitemShut {NoStop}%
\bibitem [{\citenamefont {Swingle}(2012{\natexlab{a}})}]{Swin3}%
  \BibitemOpen
  \bibfield  {author} {\bibinfo {author} {\bibfnamefont {B.}~\bibnamefont
  {Swingle}},\ }\bibfield  {title} {\enquote {\bibinfo {title} {Conformal field
  theory approach to {Fermi} liquids and other highly entangled states},}\
  }\href@noop {} {\bibfield  {journal} {\bibinfo  {journal} {Phys. Rev. B}\
  }\textbf {\bibinfo {volume} {86}},\ \bibinfo {pages} {035116} (\bibinfo
  {year} {2012}{\natexlab{a}})}\BibitemShut {NoStop}%
\bibitem [{\citenamefont {Swingle}(2012{\natexlab{b}})}]{Swin2}%
  \BibitemOpen
  \bibfield  {author} {\bibinfo {author} {\bibfnamefont {B.}~\bibnamefont
  {Swingle}},\ }\bibfield  {title} {\enquote {\bibinfo {title} {R\'enyi
  entropy, mutual information, and fluctuation properties of {Fermi}
  liquids},}\ }\href@noop {} {\bibfield  {journal} {\bibinfo  {journal} {Phys.
  Rev. B}\ }\textbf {\bibinfo {volume} {86}},\ \bibinfo {pages} {045109}
  (\bibinfo {year} {2012}{\natexlab{b}})}\BibitemShut {NoStop}%
\bibitem [{\citenamefont {Balian}(1992)}]{Ba}%
  \BibitemOpen
  \bibfield  {author} {\bibinfo {author} {\bibfnamefont {R.}~\bibnamefont
  {Balian}},\ }\href@noop {} {\emph {\bibinfo {title} {From Microphysics to
  Macrophysics, Vol. II}}}\ (\bibinfo  {publisher} {Springer--Verlag, Berlin},\
  \bibinfo {year} {1992})\BibitemShut {NoStop}%
\bibitem [{\citenamefont {Sobolev}(2015)}]{Sob3}%
  \BibitemOpen
  \bibfield  {author} {\bibinfo {author} {\bibfnamefont {A.~V.}\ \bibnamefont
  {Sobolev}},\ }\bibfield  {title} {\enquote {\bibinfo {title} {{Wiener--Hopf}
  operators in higher dimensions: the {Widom} conjecture for piece-wise smooth
  domains},}\ }\href@noop {} {\bibfield  {journal} {\bibinfo  {journal}
  {Integr. Equ. Oper. Theory}\ }\textbf {\bibinfo {volume} {81}},\ \bibinfo
  {pages} {435--449} (\bibinfo {year} {2015})},\ \Eprint
  {http://arxiv.org/abs/or arXiv: 1312.1835v2 [math.SP] (2014)} {or arXiv:
  1312.1835v2 [math.SP] (2014)} \BibitemShut {NoStop}%
\bibitem [{\citenamefont {Vidal}\ and\ \citenamefont {Werner}(2002)}]{VW}%
  \BibitemOpen
  \bibfield  {author} {\bibinfo {author} {\bibfnamefont {G.}~\bibnamefont
  {Vidal}}\ and\ \bibinfo {author} {\bibfnamefont {R.~F.}\ \bibnamefont
  {Werner}},\ }\bibfield  {title} {\enquote {\bibinfo {title} {Computable
  measure of entanglement},}\ }\href@noop {} {\bibfield  {journal} {\bibinfo
  {journal} {Phys. Rev. A}\ }\textbf {\bibinfo {volume} {65}},\ \bibinfo
  {pages} {032314} (\bibinfo {year} {2002})}\BibitemShut {NoStop}%
\bibitem [{\citenamefont {Plenio}(2005)}]{Plenio}%
  \BibitemOpen
  \bibfield  {author} {\bibinfo {author} {\bibfnamefont {M.~B.}\ \bibnamefont
  {Plenio}},\ }\bibfield  {title} {\enquote {\bibinfo {title} {Logarithmic
  negativity: A full entanglement monotone that is not convex},}\ }\href@noop
  {} {\bibfield  {journal} {\bibinfo  {journal} {Phys. Rev. Lett.}\ }\textbf
  {\bibinfo {volume} {95}},\ \bibinfo {pages} {090503} (\bibinfo {year}
  {2005})}\BibitemShut {NoStop}%
\bibitem [{\citenamefont {Gardner}(2006)}]{Gard}%
  \BibitemOpen
  \bibfield  {author} {\bibinfo {author} {\bibfnamefont {R.~J.}\ \bibnamefont
  {Gardner}},\ }\href@noop {} {\emph {\bibinfo {title} {Geometric
  Tomography}}}\ (\bibinfo  {publisher} {Cambridge University Press, 2nd.
  edition},\ \bibinfo {address} {Cambridge},\ \bibinfo {year}
  {2006})\BibitemShut {NoStop}%
\bibitem [{\citenamefont {Sobolev}(2014)}]{Sob2}%
  \BibitemOpen
  \bibfield  {author} {\bibinfo {author} {\bibfnamefont {A.~V.}\ \bibnamefont
  {Sobolev}},\ }\bibfield  {title} {\enquote {\bibinfo {title} {On the
  {Schatten--von Neumann} properties of some pseudo-differential operators},}\
  }\href@noop {} {\bibfield  {journal} {\bibinfo  {journal} {J. Funct. Anal.}\
  }\textbf {\bibinfo {volume} {226}},\ \bibinfo {pages} {5886--5911} (\bibinfo
  {year} {2014})},\ \Eprint {http://arxiv.org/abs/or arXiv: 1310.2083v2
  [math.SP] (2013)} {or arXiv: 1310.2083v2 [math.SP] (2013)} \BibitemShut
  {NoStop}%
\bibitem [{\citenamefont {Landau}\ and\ \citenamefont {Widom}(1980)}]{LW}%
  \BibitemOpen
  \bibfield  {author} {\bibinfo {author} {\bibfnamefont {H.~J.}\ \bibnamefont
  {Landau}}\ and\ \bibinfo {author} {\bibfnamefont {H.}~\bibnamefont {Widom}},\
  }\bibfield  {title} {\enquote {\bibinfo {title} {The eigenvalue distribution
  of time and frequency limiting},}\ }\href@noop {} {\bibfield  {journal}
  {\bibinfo  {journal} {J. Math. Anal. Appl.}\ }\textbf {\bibinfo {volume}
  {77}},\ \bibinfo {pages} {469--481} (\bibinfo {year} {1980})}\BibitemShut
  {NoStop}%
\bibitem [{\citenamefont {Sobolev}(2013)}]{Sob1}%
  \BibitemOpen
  \bibfield  {author} {\bibinfo {author} {\bibfnamefont {A.~V.}\ \bibnamefont
  {Sobolev}},\ }\bibfield  {title} {\enquote {\bibinfo {title}
  {Pseudo-differential operators with discontinuous symbols: Widom's
  conjecture},}\ }\href@noop {} {\bibfield  {journal} {\bibinfo  {journal}
  {Mem. Am. Math. Soc.}\ }\textbf {\bibinfo {volume} {222}},\ \bibinfo {pages}
  {1043} (\bibinfo {year} {2013})},\ \Eprint {http://arxiv.org/abs/or arXiv:
  1004.2576v2 [math.SP] (2011)} {or arXiv: 1004.2576v2 [math.SP] (2011)}
  \BibitemShut {NoStop}%
\bibitem [{\citenamefont {Widom}(1982)}]{Widom_82}%
  \BibitemOpen
  \bibfield  {author} {\bibinfo {author} {\bibfnamefont {H.}~\bibnamefont
  {Widom}},\ }\bibfield  {title} {\enquote {\bibinfo {title} {On a class of
  integral operators with discontinuous symbol},}\ }\href@noop {} {\bibfield
  {journal} {\bibinfo  {journal} {Toeplitz Centennial (Tel Aviv, 1981),
  Operator Theory: Adv. Appl.}\ }\textbf {\bibinfo {volume} {4}},\ \bibinfo
  {pages} {477--500} (\bibinfo {year} {1982})}\BibitemShut {NoStop}%
\bibitem [{lee()}]{leer}%
  \BibitemOpen
  \href@noop {} {\ }\BibitemShut {NoStop}%
\bibitem [{\citenamefont {Simon}(2005)}]{Simon}%
  \BibitemOpen
  \bibfield  {author} {\bibinfo {author} {\bibfnamefont {B.}~\bibnamefont
  {Simon}},\ }\href@noop {} {\emph {\bibinfo {title} {Trace Ideals and Their
  Applications, 2nd edition}}}\ (\bibinfo  {publisher} {American Mathematical
  Society},\ \bibinfo {address} {Providence, RI},\ \bibinfo {year}
  {2005})\BibitemShut {NoStop}%
\bibitem [{\citenamefont {Birman}\ and\ \citenamefont {Solomyak}(1987)}]{BS}%
  \BibitemOpen
  \bibfield  {author} {\bibinfo {author} {\bibfnamefont {M.~S.}\ \bibnamefont
  {Birman}}\ and\ \bibinfo {author} {\bibfnamefont {M.~Z.}\ \bibnamefont
  {Solomyak}},\ }\href@noop {} {\emph {\bibinfo {title} {Spectral Theory of
  Self-adjoint Operators in Hilbert Space}}}\ (\bibinfo  {publisher} {D. Reidel
  Publishing Company, Dordrecht, Holland},\ \bibinfo {year} {1987})\BibitemShut
  {NoStop}%
\end{thebibliography}%

\end{document}